\documentclass[sigconf]{acmart}
\AtBeginDocument{%
  }

\usepackage[normalem]{ulem}
\usepackage{cancel}

\copyrightyear{2026}
\acmYear{2026}
\setcopyright{cc}
\setcctype{by}
\acmConference[CHI '26]{Proceedings of the 2026 CHI Conference on Human Factors in Computing Systems}{April 13--17, 2026}{Barcelona, Spain}
\acmBooktitle{Proceedings of the 2026 CHI Conference on Human Factors in Computing Systems (CHI '26), April 13--17, 2026, Barcelona, Spain}
\acmPrice{}
\acmDOI{10.1145/3772318.3791132}
\acmISBN{979-8-4007-2278-3/2026/04}
\begin{document}

\title{Reimagining Data Work: Participatory Annotation Workshops as Feminist Practice}

\newcommand{\revise}[1]{\textcolor{black}{#1}}
\newcommand{\final}[1]{\textcolor{black}{#1}}

\newcommand{\todo}[1]{\textcolor{red}{\textbf{TODO: #1}}}
\newcommand{\yujia}[1]{\textcolor{orange}{\textbf{yujia: #1}}}
\newcommand{\isa}[1]{\textcolor{purple}{\textbf{Isa: #1}}}
\newcommand{\harini}[1]{\textcolor{brown}{\textbf{Harini: #1}}}
\newcommand{\helena}[1]{\textcolor{green}{\textbf{Helena: #1}}}

\author{Yujia Gao}
\email{yujia_gao@brown.edu}
\affiliation{%
  \institution{Brown University}
  \city{Providence}
  \state{RI}
  \country{USA}
}
\author{Isadora Cruxên}
\email{i.cruxen@qmul.ac.uk}
\affiliation{%
  \institution{Queen Mary University of London}
  \city{London}
  \country{UK}
}

\author{Helena Suárez Val}
\email{ladelentes@gmail.com}
\affiliation{%
  \institution{Feminicidio Uruguay}
  \city{Ciudad de la Costa}
  \country{Uruguay}
}

\author{Alessandra Jungs de Almeida}
\email{jungsdea@mit.edu}
\affiliation{%
  \institution{Data + Feminism Lab, MIT}
  \city{Cambridge}
  \state{MA}
  \country{USA}
}

\author{Catherine D’Ignazio}
\email{digazio@mit.edu}
\affiliation{%
  \institution{Data + Feminism Lab, MIT}
  \city{Cambridge}
  \state{MA}
  \country{USA}
}

\author{Harini Suresh}
\email{harini_suresh@brown.edu}
\affiliation{%
  \institution{Brown University}
  \city{Providence}
  \state{RI}
  \country{USA}
}


\begin{abstract}
AI systems depend on the invisible and undervalued labor of data workers, who are often treated as interchangeable units rather than collaborators with meaningful expertise. 
Critical scholars and practitioners have proposed alternative principles for data work, but few empirical studies examine how to enact them in practice. 
This paper bridges this gap through a case study of multilingual, iterative, and participatory data annotation processes with journalists and activists focused on news narratives of gender-related violence. We offer two methodological contributions. 
First, we demonstrate how workshops rooted in feminist epistemology can foster dialogue, build community, and disrupt knowledge hierarchies in data annotation. Second, drawing insights from practice, we deepen analysis of existing feminist and participatory principles. We show that prioritizing context and pluralism in practice may require ``bounding'' context and working towards what we describe as a ``tactical consensus.'' We also explore tensions around materially acknowledging labor while resisting transactional researcher-participant dynamics. Through this work, we contribute to growing efforts to reimagine data and AI development as relational and political spaces for understanding difference, enacting care, and building solidarity across shared struggles.

\end{abstract}



\renewcommand{\shortauthors}{Gao et al.}



\maketitle

\textit{Content warning: Figure 1 in this paper displays a news article describing domestic violence and feminicide.}
\section{Introduction}
Large-scale artificial intelligence (AI) research and development is currently concentrated in the hands of large corporations \cite{verdegemDismantlingAICapitalism2024}, even as it increasingly relies on a vast, largely invisible ``human infrastructure'' of data annotators \cite{mateescu2019ai}. Annotators label social media text used in content moderation systems \cite{chen2014laborers}, interpret facial expressions to train diagnostic tools \cite{catanzariti_seeing_2023}, rank large language model (LLM) outputs to make them more ``helpful and honest'' \cite{bai2022training}, and many other tasks \cite{wang2022whose}. Their annotations fundamentally shape how we make sense of data, and how AI systems represent the world.  

Despite their indispensable role in data production, annotators generally have little control over the annotation process. Moreover, data are typically removed from their context during annotation. Rather than collaborating with individuals who might have relevant lived experience or cultural knowledge, annotation is often outsourced to anonymous workers via digital platforms that replicate existing power structures through low wages, job precarity, and worker surveillance \cite{miceli2022data}. This setup reflects a broader ethos of data work: that it is “rote” work that can be delegated, rather than a complex process of sense-making that requires deep expertise, deliberation, and cultural sensitivity \cite{feinberg2022everyday}. As it trickles down into concrete datasets and AI products, this ethos perpetuates extractive and colonial structures in AI development \cite{muldoon2024feeding, ricaurte2019data}. 

Work across feminist Human-Computer Interaction (HCI) and participatory AI has argued for AI development practices that consider context, make labor visible, distribute decision-making power, and center the perspectives of historically marginalized communities, critically examining existing practices and proposing frameworks or recommendations to guide future work \cite{Klein2024DF, ricaurte2022introduction,garcia2022no,d2020toward,miceli2022studying,Leavy2021EthicalDC,grayFeministDataEthics2021, corbett2023power}. \revise{ However, while a growing body of scholarship articulates guiding principles or calls for alternative approaches, few published examples examine and reflect on the frictions of enacting critical, participatory and/or feminist approaches in practice for AI development \cite{ciolfifeliceDoingFeministWork2025, sureshIntersectionalFeministParticipatory2022,data_workers_inquiry}. }


\revise{To address this gap, this paper bridges conceptual frameworks and concrete practice in AI development through a case study of participatory data annotation.} We describe the design and implementation of a multilingual, collaborative, and iterative data annotation process focused on news coverage of gender-related violence. \revise{Drawing on empirical data from the participatory data annotation process and our own ongoing critical reflections, our case study offers two key methodological contributions. First, it shows how workshops can be intentionally structured to move abstract commitments into enacted practices that engage diverse, contextual perspectives in AI development, while also generating insights into how AI applications might be reshaped by such practices. We mobilized the workshop as an intentional feminist space to facilitate collective dialogue and collaboration in a relational way---rather than individualized, isolated labor---and to enable creative experimentation. Second, by deriving empirical insights from practice, our work deepens the discussion of existing normative principles for alternative approaches to AI. Our study shows, for example, that embracing plural perspectives \cite{DIgnazio2022DataF} requires not simply considering context but also intentionally ``bounding'' it. This brings into view tensions surrounding the role of standardization in data annotation, prompting reflection on what we describe as \textit{tactical consensus} (Section \ref{sec:context}). We also highlight the importance of creating what we call \textit{multiple depths of dialogue} for balancing meaningful participation with its affective burdens (Section \ref{sec:dialogue}), and explore tensions around materially acknowledging labor while resisting transactional researcher-participant dynamics (Section \ref{sec:labor})}. These insights help to bridge political economy critiques of data annotation work with participatory approaches to AI development, extending critical and feminist HCI thinking on AI.

\revise{This work provides a methodological contribution to scholars, technologists, and activists seeking alternatives to extractive, cor-porate-led AI development practices. We place a focus on data annotation---a step in AI development where, as much work has described, the processes and people involved are often hidden and undervalued \cite{wang2022whose,toxtli2021quantifying,sambasivan2021everyone,gray2019ghost,Miceli2020BetweenSA,miceli2022data}---to interrogate how critical frameworks for AI development can be practically enacted.} The paper is organized as follows. In Section \ref{sec:related-work}, we review related work to situate our \revise{methodological proposal} in relation to existing literature on data annotation and feminist and participatory HCI/AI, as well as longer-standing traditions of feminist dialogue and organizing. In Section \ref{sec:method}, we describe the goals, iterative design, and rationales for \revise{our methodological choices in organizing a series of \revise{collaborative data annotation} workshops to explore possibilities for} technological tools to support responsible and feminist reporting of gender-related violence.
\revise{In Section \ref{sec:reflections}, drawing on a qualitative analysis of empirical data, including workshop recordings, feedback surveys, and brainstorming boards, we distill methodological insights and enduring tensions around considering context, fostering collective dialogue, acknowledging labor, and centering care in practice. }
In Section \ref{sec:discussion}, we conclude by discussing how our experiences open up avenues for future efforts to reimagine data and AI work in diverse contexts. \revise{Building on \citet{ciolfifeliceDoingFeministWork2025}'s provocation around ``building AI otherwise,'' we reflect on how initiating an ``AI development'' process served as a heuristic device, creating space for us, together with participants, to critically and collectively engage with the problem of harmful media reporting, explore alternative data annotation practices that center care and context, and grapple with the broader question of if---and under what conditions---AI should play any role in this domain.}


\section{Related Work}\label{sec:related-work}

\subsection{Data Annotation Practices in AI}

Data annotation is a critical component of AI development. While most closely associated with supervised learning approaches, data annotation also underpins many contemporary post-training processes. For instance, LLMs often rely on annotated data for fine-tuning and alignment through methods such as reinforcement learning with human feedback (RLHF) \cite{casper2023open}. Beyond training, human annotations are commonly used to evaluate systems, such as through ranking or rating model outputs. Data annotation typically involves formulating a specific task, developing a labeling scheme or taxonomy to represent “ground truth,” and applying this to (or ``labeling'') the data. Example tasks \revise{include} classifying images of handwritten digits \cite{Deng2012TheMD}, identifying facial microexpressions from images \cite{catanzariti_seeing_2023}, rating the harmfulness of generated text or images \cite{Wang2025JustAS}, and others. Data annotation is often crowdsourced, resulting in tasks being performed independently by large numbers of \revise{(often unpaid or low-paid) workers}. This paradigm is consolidated through platforms such as Amazon Mechanical Turk (MTurk)\footnote{https://www.mturk.com/} or ScaleAI\footnote{https://scale.com/} that recruit workers to annotate data at scale. 

A range of research has characterized how the \revise{infrastructure} of data work creates and maintains power imbalances \cite{miceli2022data,Miceli2020BetweenSA,Miceli2022Documenting,sambasivan2021everyone,posada2022embedded,rifat2024data}. Categorization schemes and instructions for data annotation are primarily developed by small groups of researchers and AI practitioners. This hierarchical structure means such instructions typically hold strong authority, and are accepted as ``correct'' or ``indisputable'' \cite{Miceli2020BetweenSA, miceli2022data}. The global economic structure of data work further reflects stark hierarchies in knowledge production \cite{muldoon2024feeding}. Whereas research and technical development originate mostly in the Global North, data annotation work is usually outsourced to low-income countries in the Global Majority \cite{Daz2022CrowdWorkSheetsAF, Sambasivan2021India,miceli2022data}. At these sites, annotators’ work is often underpaid and lacks legal protection \cite{gray2019ghost, tubaro2025digital, Kotaro2018Turk}. These conditions reinforce colonial histories, with annotators viewed as distant, interchangeable units of labor, rather than co-contributors with valuable expertise and a stake in the technology being produced \cite{Irani2013TurkopticonIW, Rothschild2024ThePW}. Language also plays a central role in these hierarchies: AI’s reliance on English privileges Western epistemologies while marginalizing others. As recent studies show, English training datasets are often treated as “universal” and AI systems perform poorly when processing other languages \cite{Moorosi2024} or representing non-Western contexts \cite{Hall2024TowardsGI}. 

Recent literature has challenged these power hierarchies. Against the idea that annotators are interchangeable units, research shows that annotators’ identity, social position, and cultural backgrounds shape how they apply knowledge and annotate \cite{Prabhakaran2021OnRA, Daz2022CrowdWorkSheetsAF}, and that annotator disagreements often reflect nuances that were missed in task formulation or taxonomy, particularly for tasks requiring an understanding of social and cultural contexts \cite{Plank2014LinguisticallyDO, Smart2024DisciplineAL}. The Data Workers' Inquiry project emphasizes the situated expertise of data workers, supporting them in formulating research questions and conducting inquiries around the material and political conditions of their work \cite{data_workers_inquiry}. Other work has proposed or implemented ways to improve data annotation through documentation \cite{Daz2022CrowdWorkSheetsAF, Miceli2022Documenting}, developing annotation tools and systems \cite{Kuo2024WikibenchCD}, collaborating with domain experts \cite{Patton2020ContextualAO, maConceptualQuestionsDeveloping2023}, improving employment structures \cite{wang2022whose}, and respecting annotators' qualifications and expertise \cite{wang2022whose, Rothschild2022Data}. Particularly relevant to our case study is the Contextual Analysis of Social Media (CASM) approach developed by \citet{Patton2020ContextualAO}. Their research team employed a collaborative method that combined academic expertise and the expertise of community members with lived experiences relevant to the context of their work. This approach allowed them to understand subtle contextual meanings in gang-related Twitter posts involving texts, images, and emojis that may otherwise have been overlooked. 

Our work builds on these efforts to shift exploitative data annotation practices. In what follows, we review literature on feminist and participatory design and AI development that informed the design principles and goals of our collaborative data annotation processes. \revise{We highlight three principles drawn from this existing scholarship: \textit{acknowledging labor, considering context and pluralism, and centering care}. Expanding on these, we articulate a fourth principle that grounded our methodology: \textit{fostering collective dialogue.}}


\subsection{Feminist and Participatory AI}
Participatory approaches to AI stem from multiple disciplines, including participatory design (PD) \cite{Gregory2003ScandinavianAT, Simonsen2012RoutledgeIH} and participatory action research (PAR) \cite{Baum2006ParticipatoryAR}. Though they take many forms and have varied motivations, participatory AI efforts broadly aim to involve a wide range of relevant stakeholders into deliberation and design processes \cite{delgado2023participatory}. In data science and AI, participatory approaches have been increasingly recognized as a way to challenge default power imbalances in technological development \revise{through shifting decision-making power or system ownership to those who are marginalized by structural inequalities} \cite{corbett2023power,birhane2022power}. At the same time, participatory scholarship warns against the risk of “participation-washing,” or the tokenistic involvement of stakeholders without meaningful sharing of power—a concern that is particularly salient in current AI development given its increasing centralization and corporate control \cite{arnstein1969ladder,sloane2022participation,suresh2024participation}. Some case studies describe participatory approaches in AI development or public engagement that are mindful of this risk \cite{sureshIntersectionalFeministParticipatory2022,nekoto2020participatory,nucera2018people,katell2020toward}. Our work builds on these approaches, and is particularly focused on the unique knowledge production processes involved in data annotation.


Our approach to designing alternative, participatory data annotation practices is grounded in intersectional feminist theory and ethics of care. Intersectional feminist theory, originating from Black feminist thought, attends to the relational and embodied experiences of those experiencing multiple, interlocking forms of oppression \cite{crenshaw1989demarginalizing, combahee1977black, collins2019intersectionality}. Drawing from this, Data Feminism proposes a set of principles for thinking about and working with data in ways that take structural inequalities seriously \cite{DIgnazio2022DataF}. Three of these principles directly guided our work: ``make labor visible,'' ``consider context,'' and ``embrace pluralism.'' 
As we will discuss further, in our research, we sought to \textit{acknowledge the labor}---intellectual, emotional, and embodied---associated with data annotation, and to prioritize processes that respect the personhood, rights, and well-being of annotators as collaborators. We also sought to \textit{consider context and pluralism} by engaging participants from diverse and relevant cultural and geographic backgrounds, as well as with varying expertise on the topics we work with, gender-related violence and feminicide. 

Alongside these principles, our approach draws from feminist ethics of care \cite{de2017matters, carecollective2020care} to \textit{center care} in data annotation work. Care ethics—while not a uniform theory—is broadly founded on the understanding that people exist in “mutually interconnected, interdependent, and often unequal relations with each other,” \cite{hankivsky2014rethinking} elevating the importance of relationality, compassion, and community care. This stands in contrast to epistemological theories that prioritize objective, individualistic, or unemotional ways of thinking \cite{tronto1998ethic}. Intersectional critiques of care ethics have led to conceptualizations that view self- and collective care as serving an important political (not only interpersonal) role within intersecting structures of power \cite{graham2007ethics}. 

In relation to data and AI, conceptual scholarship has explored how care ethics provides a new lens to understand and respond to datafication \cite{zakharovaCarefulDataStudies2024}, to reflect on research practices \cite{Luka2018ReframingBD}, and to motivate interventions into the AI pipeline, such as increased transparency of technical artifacts or diversity on technical teams \cite{grayFeministDataEthics2021}. 
Expanding on the role of care for societal change, \citet{DignazioCF} conceptualizes ``restorative/transformative data science,'' where data production contributes to the twin goals of (1) restoration —redressing the immediate, shorter-term consequences of violence, engaging in healing for impacted communities, and participating in efforts to restore vitality, dignity and rights; and (2) transformation—working towards the longer-term dismantling of the structural conditions that perpetuate inequality. \revise{Other frameworks draw from library studies \cite{Jo2019LessonsFA} and critical race theory \cite{Leavy2021EthicalDC} to guide alternative approaches to data curation.} 

\revise{While prior work offers valuable conceptual aims, few examples examine the frictions of enacting these critical, participatory, and feminist approaches in practice for AI development \cite{ciolfifeliceDoingFeministWork2025,sureshIntersectionalFeministParticipatory2022}. With respect to data annotation,} emerging studies experiment with care- and worker-centered annotation, such as building data workers' AI literacy \cite{DiSalvo2024Worker} and prioritizing annotators' well-being in labeling sensitive content \cite{lameiroTIDEsTransgenderNonbinary2025}. \revise{Our work contributes to bridging conceptual frameworks and concrete practices, drawing methodological insights from the implementation of participatory, collaborative, and care-centered data annotation workshops.} The following subsection expands on this methodology.


\subsection{Workshops and Collective Dialogue}

Two broad veins of academic work and organizing practice informed our methodological approach. The first draws from PD \cite{shaowenUtopias2018}, transition design \cite{juri2021transition}, design justice \cite{Sasha2020DesignJC} and feminist HCI \cite{Bardzell2010FeministHT}, wherein researchers and grassroots communities create collaborative spaces to explore or co-design technology, data or AI tools through workshops \cite{rosner2016out}, hackathons \cite{hopeHackathonsParticipatoryDesign2019, DIgnazio2020ThePI}, and other spaces \cite{fox2015hackerspace}. For example, \citet{ConundrumPoliceOfficerinvolved} engaged local communities in a ``counterdata action,'' an event to collectively analyze civic data and build counter-narratives about police brutality. Other HCI work used design workshops to engage specific communities, such as youths \cite{Delcourt2022InnovatingNO, Vacca2017BiculturalET}, older adults \cite{Harrington2019EngagingLA}, or people with aphasia \cite{Galliers2012WordsAN}. Reflections on the personal and political impacts of these efforts illustrate their promise to serve as spaces for consciousness-raising and critical literacy \cite{DIgnazio2020ThePI,DiSalvo2024Worker}.

The second vein draws on historically situated feminist mobilization and organizing. Feminist meetings have long served as cornerstones for political articulation and collective action aimed at structural transformation. These gatherings take many forms: from women’s suffrage conventions in the US, to feminist consciousness-raising groups that enabled women to share lived experiences and resist structural oppression \cite{kaba2017trying,hanisch2010women}, to Latin American “encuentros” against violence and for reproductive justice \cite{lvarez2003EncounteringLA}. Within the Latin American feminist tech movement, in particular, workshops are recognized as a powerful mode of gathering, valued for their ability to foster active participation, collaboration, creativity, learning, and emotional support \cite{pedraza2019feminist_technopolitics}. 

\revise{Building on these two veins of work, we articulated \textit{fostering collective dialogue} as a fourth key principle that guided our approach.} Our use of the term dialogue draws on the work of Brazilian educator Paulo Freire, which emphasizes the social and transformative character of knowledge production and learning \cite{freire1970pedagogy, freire1980extensao}. For Freire, dialogue implies communication and, therefore, reciprocity. It entails both critical reflection and action. This perspective resists the notion that knowledge can simply be “extended” or transferred from one person to another; instead, learning happens through a communicative process. Dialogue is then an active process that enables people to understand themselves as subjects, that is, as historical agents of transformation. 

\revise{Our methodological contribution illustrates the design of collaborative feminist data annotation workshops that gather and engage a community of practice in collective dialogue, serving as a space for collective and personal critical reflection.} \revise{For us, a crucial dimension of workshops as a methodology is the opportunity they create for horizontal dialogue between participants, countering} the typically isolated and atomized nature of data annotation pipelines, and creating instead a mode of AI development that fosters criticality and collectivity.

\section{Participatory Data Annotation in Practice: Co-creating an Annotated Dataset on Feminicide News}\label{sec:method}

We detail a case study of building a multilingual, collaboratively annotated dataset of news coverage of feminicide, as the beginning of a broader project exploring the possibility for AI tools to support responsible news reporting on gender-related violence. The dataset was developed through collaborative data annotation workshops combined with continued, asynchronous annotation. We describe the background, design, and implementation of our process below. 

\subsubsection{Research team}
Before proceeding with the case study narrative, it is relevant to introduce the ``we'' writing this paper. We are six researchers working interdisciplinarily from computer science, political economy, and critical and feminist data studies. 
All but the first author have worked together for the past five years in the Data Against Feminicide (DAF) project\footnote{https://datoscontrafeminicidio.net/en/home/}, which seeks to understand and support activist data practices and foster a transnational research-activist network focused on feminicide and gender-related violence.
The third author has carried out feminicide monitoring via news media in Uruguay for over 10 years.
The first author joined the team as a member of a US-based academic lab founded by the last author, which continues to work in partnership with DAF. 

The present project is a continuation of this ongoing collaboration, \revise{which has framed some methodological decisions}, such as adopting the DAF working languages (English, Spanish, and Portuguese) in the first workshop, and provided a strong starting base, including existing relationships with a large community of activists, journalists, academics, and civil society groups who work with data on feminicide, and access to the DAF tools for compiling news articles. Importantly, one aspect of the DAF project has been the co-design of technological interventions to support community partners. In collaboration with activists, the project developed two AI-based tools that help activists filter through news reports of violent events and record information about feminicide cases. This prior collaborative work provided the impetus for this study, as both within our team and amongst the activists we worked with there was interest in exploring whether AI-based tools developed collaboratively and from a feminist lens could also help to address media narratives of feminicide. Echoing \citet{ciolfifeliceDoingFeministWork2025}, we wanted to explore ``when and to what extent AI
techniques can be used to leverage feminist and social justice causes.''

\subsection{Background and Problem Statement}

News media are crucial for disseminating information and shaping public understandings of social issues. Specifically, news coverage of gender-related violence and feminicide has been widely critiqued for reproducing sexist stereotypes, distancing readers from the issue, blaming victims for their own murders, failing to cover violence against marginalized communities, and framing cases as isolated incidents rather than as part of a broader and systemic problem \cite{aldreteFramingFemicideNews, Casados2018, Oliveira2021}. Scholars have shown that these patriarchal narratives not only maintain and reinforce structural oppressions but also exacerbate state negligence and public inaction \cite{DignazioCF}. Despite these challenges, feminist data activists \revise{largely} rely on news to produce feminicide data, especially when the State neglects to do so \cite{DIgnazio2022FeminicideAC, young2022justice}. 

Shifting harmful narratives around gender-related violence is thus a crucial area of work. 
The current project started by asking: Could technology support activists and journalists in achieving more responsible news coverage? \revise{We wanted to keep the project outcomes broad so participants could help shape its direction, but we also wanted to explore, as mentioned earlier, whether and how AI might fit into this context.} 
For example, could an AI model identify harmful reporting practices, such as victim-blaming, and signal to more constructive ones, such as situating incidents within a broader \revise{pattern of structural gendered violence}? Could a tool with this capability serve a pedagogical role for journalists and editors, or be a way for activists to hold news organizations accountable? As we considered these questions, we also wondered: Could the development of such a tool become a space to think critically about the role of AI in feminist activism and media production, and to explore alternative visions of AI development? To approach these questions, we decided to initiate a co-design process with \revise{feminist activists and journalists}, whereby we hoped to envision potential tools, as well as identify aspects in which \revise{AI---or automation more generally---}would \textit{not} be helpful or desirable. A first step towards experimenting with these ideas was to create a dataset representing both harmful language and stereotypes as well as constructive elements in media coverage, which could be used \revise{as a basis for ideating, building, or evaluating future tools.} 

Collecting and annotating such a dataset is not straightforward. How should relevant harmful and constructive practices be identified? Reporting practices and societal narratives around gender-related violence are complex and culturally specific. We can partially draw from existing best practice guides at local, national, and international scales \cite{fernandez2023guia, ifj_guidelines_vaw, un_guidelines_reporting}, but doing so has its own challenges. Local reporting guidelines can conflict with each other and with guidelines produced by international organizations like the International Federation of Journalists \cite{ifj_guidelines_vaw} or UN Women \cite{un_guidelines_reporting}). For instance, there is no consensus on whether including victims’ names in news reporting constitutes a good or bad practice: some media organizations choose to do so, arguing that this humanizes victims and honors their memory. In contrast, other contexts avoid naming for cultural or privacy reasons: 
In Korea, for example, news media typically refrain from publishing victims’ names in the news to avoid violating the privacy of victims and their families. For \revise{the Indigenous Apache peoples and the Aboriginal groups in the territory of what is today Australia}, naming the dead constitutes a cultural taboo\revise{, and its violation is considered profoundly disrespectful} \cite{Palmer2008Supernatural}. 
Recognizing these context-specific tensions, we organized a series of collaborative workshops to discuss and co-develop a taxonomy of reporting practices and a participatory annotation process, as well as to collectively shape the future of the project. 

\subsection{From Design Principles to Iterative Design Practice }
In this section, we describe translating the design principles articulated in Section \ref{sec:related-work} into methodological choices and practices for collaborative data annotation. Between December 2024 to October 2025, we organized a series of engagements: an introductory trilingual workshop,
two single-language workshops, and a longer phase of asynchronous annotation. In total, 79 people from 35 countries with diverse expertise on gender-related issues and/or journalism participated in the process. The project was reviewed by the Institutional Review Board at the first author’s institution, and was deemed exempt from full review. Grounded in critical HCI work emphasizing reflexivity \cite{rosner2016out}, our process was deliberately iterative: feedback from each stage shaped the redesign and organization of the following stages. We begin by explaining how the participatory process unfolded, interweaving our observations and reflections to show how they informed our evolving approach. Then, in Section \ref{sec:reflections}, we systematize our insights on the challenges encountered and the responses we devised while trying to enact our design principles in practice.

\subsubsection{Introductory Workshop}\label{sec:intro-workshop}

Our process began with a trilingual online workshop held in December 2024 with participation from 50 activists, journalists, and researchers. Participants were recruited via email outreach to DAF's existing transnational research-activist network and via targeted invitation to journalist groups and organizations. The community we gathered makes up a ``community of practice'' \cite{wenger2015introduction}, where participants share an interest and engagement with feminicide and data activism, rather than sharing a geographic or cultural identity. The majority of participants (60\%) were Spanish-speakers, followed by 32\% English- and 8\% Portuguese-speakers. This distribution reflects longstanding transnational efforts to build and strengthen a community of practice across the Americas and beyond to address feminicide and gender-related violence \cite{DAF5yrReport}. 

In line with our design principles, this first workshop was designed to enable collective dialogue and community-building rather than annotation volume. To that end, it was structured in three parts: 
\begin{enumerate}
 \item \textit{Presentations}: The workshop began with brief introductions about the AI lifecycle and the role of data annotation, existing feminist AI approaches, and the project’s goals. These talks were intended to (a) situate the project’s aims within the current landscape of data work in AI development, and (b) create a shared foundation for critical engagement with AI amongst participants.
 \item \textit{Breakout Rooms}: In facilitated single-language breakout rooms with 5-10 people each, participants engaged in collaborative data annotation, where they individually read 1-3 news articles, labeled them with an initial working taxonomy, and discussed the articles and annotations as a group. We designed the smaller groups to cultivate direct relationship building amongst participants and facilitators, and to support mutual learning and collective dialogue. 
 \item \textit{Reflections and Mourning}: The workshop concluded with participants providing reflections and feedback on the annotation task on Padlet\footnote{https://padlet.com}, a collaborative bulletin board-style tool. To close, we collectively honored the victims and survivors of gender-related violence and highlighted the importance of caring for the self through a reflective breathing exercise. 
\end{enumerate}

In preparation for the workshop, the research team developed an initial taxonomy by reviewing a selection of local and international guidelines \cite{fernandez2023guia, ifj_guidelines_vaw, un_guidelines_reporting}. This process also drew on the third author's expertise and prior research 
\cite{garbayoPressUnbiasedIdentificaciona}.
In contrast to conventional data annotation pipelines, this initial taxonomy was not intended as a ground truth, but rather as a starting point for participants to build upon. \revise{Since English was the research team's working language given our different national backgrounds, }the taxonomy was developed in English \revise{and translated into Spanish and Portuguese by bilingual members of the team}, then shared with participants before the workshop to familiarize. 

\begin{figure*}[htbp]
 \centering
 \includegraphics[alt={Screenshot of Label Studio interface, shows a news article with annotations}, width=\linewidth]{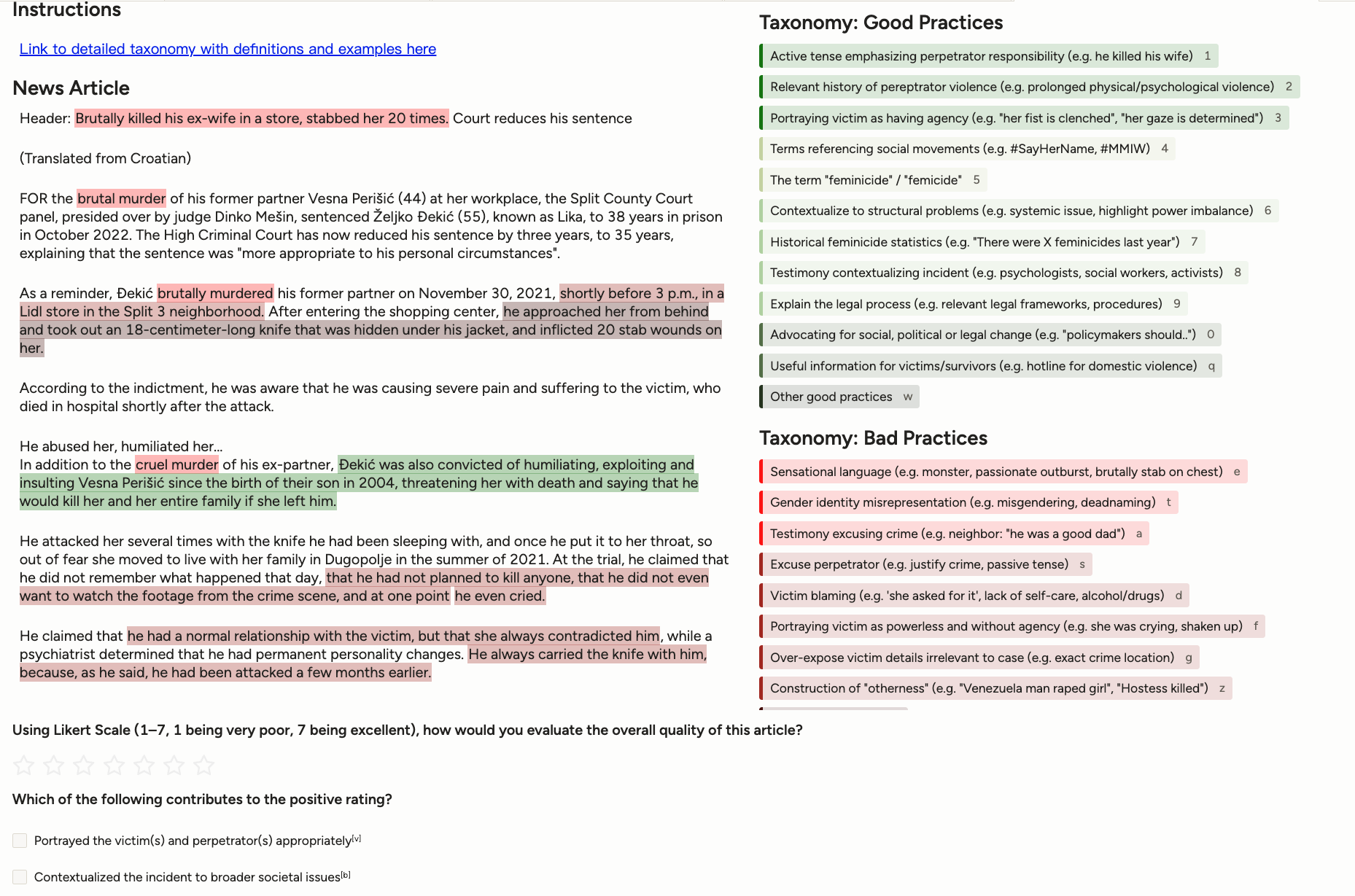} 
 \caption[An example annotation interface on Label Studio.]{An example annotation interface on Label Studio.}
 \Description{Figure 1 shows an example annotation interface in Label Studio used in our annotation processes. The left-hand side shows a news article reporting feminicides, and the right-hand side has a list of good and bad practices that participants can select to highlight sentences or phrases from the article. }
 \label{fig:interface}
\end{figure*}

To ensure meaningful engagement in the multilingual setting, we implemented live interpretation for plenary sessions and created language-specific breakout rooms moderated by facilitators fluent in the assigned language. We used Label Studio \cite{LabelStudio} as the annotation software, with the interface localized to participants’ preferred language. A selection of news articles, in all three languages, was curated from the DAF email alert system---a platform built to aid activists in detecting cases of feminicide. Figure \ref{fig:interface} shows an example annotation interface for English-speaking participants. 

All sessions were recorded with the consent of the participants. After the workshop, we conducted open coding on the breakout room transcripts and Padlet feedback to refine the task and taxonomy. For example, drawing on discussions that took place around how images in news articles often enact good/bad practices and frame the text, we expanded the annotation task to include visual content. We also added categories proposed by participants, such as the use of passive tense to excuse the perpetrator---a common practice in Spanish-language news coverage of gender-related violence. 

While the introductory workshop showed promise as a collaborative space for dialogue and knowledge co-construction, it also surfaced critical challenges that shaped our iterated workshop design. First, while the format was intended to facilitate diversity across contexts, participants reported that the multilingual translation of materials and presentations obscured some contextual nuances. For instance, the taxonomy included example phrases corresponding to each category; in translating these examples across languages, the English ``passionate outburst''---an example of sensational language---was interpreted as ``crimen pasional'' in Spanish, a loaded concept historically used to excuse or justify feminicides. 
Second, reaching consensus on a ``general'' taxonomy across linguistic and cultural contexts proved to be difficult. For example, participants diverged on whether news reports should include the exact location of the crime, reflecting different jurisdictional norms in crime documentation. 
As we considered how to proceed with the project, these insights led us to organize follow-up language-specific workshops, creating spaces for more localized dialogue.

\subsubsection{Followup Workshops}

We organized two language-specific workshops: one in Spanish in May 2025 and one in English in July 2025. 
We scheduled the Spanish-language workshop first due to strong interest in progressing the project from a larger group of Spanish speakers, and intentionally built in time between sessions to iterate on our workshop format. The Spanish and English workshops had 31 and 14 participants, respectively. 

For each language, we organized a preparatory meeting to align with participants on project goals, expected time commitments, and compensation. This was followed by a 2-hour workshop a few weeks later. 
Following a participant's suggestion, we invited all participants to contribute recent news articles from their own regions prior to the main workshop to further ground discussions on cultural and geographic contexts. During the workshop, we briefly recapped the project and reiterated the goal for participants to shape key aspects of its future. Then, the session had two key components: (1) small breakout rooms to read, annotate, and discuss news articles, with a particular focus on understanding whether the current taxonomy of reporting practices captured or missed important factors, and (2) brainstorming sessions to critically ideate potential \revise{technological tools---AI-based or otherwise---that could be developed to help support better reporting practices.} 

To cultivate deeper dialogue and create space for relationship-building, we decided again to design small breakout rooms, this time with 3-5 people. Across both workshops, we organized the breakout rooms by geographic proximity---sometimes at the country-level, sometimes broader regions---as a way to situate discussions within shared contexts. We reflect on how these groupings captured contexts differently in Section \ref{sec:context}. Based on registration volume, having smaller rooms required additional facilitators beyond the research team. Prior to the workshops, we invited some of the registered participants with relevant experience and capacity to take a more active role in the workshop as breakout room facilitators. This was an intentional choice aimed at distributing facilitation power and opening up opportunities for different levels of engagement. Facilitators received orientation material and participated in a pre-workshop meeting. Participants received \revise{USD} \$60 for engaging in the preparatory session and workshop, while those who took on facilitation roles received \revise{USD} \$100.

To involve participants in co-constructing the future of the project, we designed a brainstorming session as a key component of this round of workshops. It began with idea generation on a collaborative sticky-note board to brainstorm potential technological tools, both AI-based and otherwise, to support news production and readership (Fig. 2). Then, it transitioned into a scenario-based design exercise, where participants engaged with a speculative case study of a journalist using AI to assist with news article revision, articulating their \textit{fears}, \textit{concerns}, \textit{hopes}, and \textit{strategies} around such a tool in four quadrants of the broad. The prompt was intentionally designed to connect participants with the affective dimensions of entanglements with AI, helping to ``elevate emotion'' in data science \cite{DIgnazio2022DataF}.

\begin{figure*}[htpb]
 \centering
 \includegraphics[alt={Screenshot of digital board with sticky notes}, width=1\linewidth]{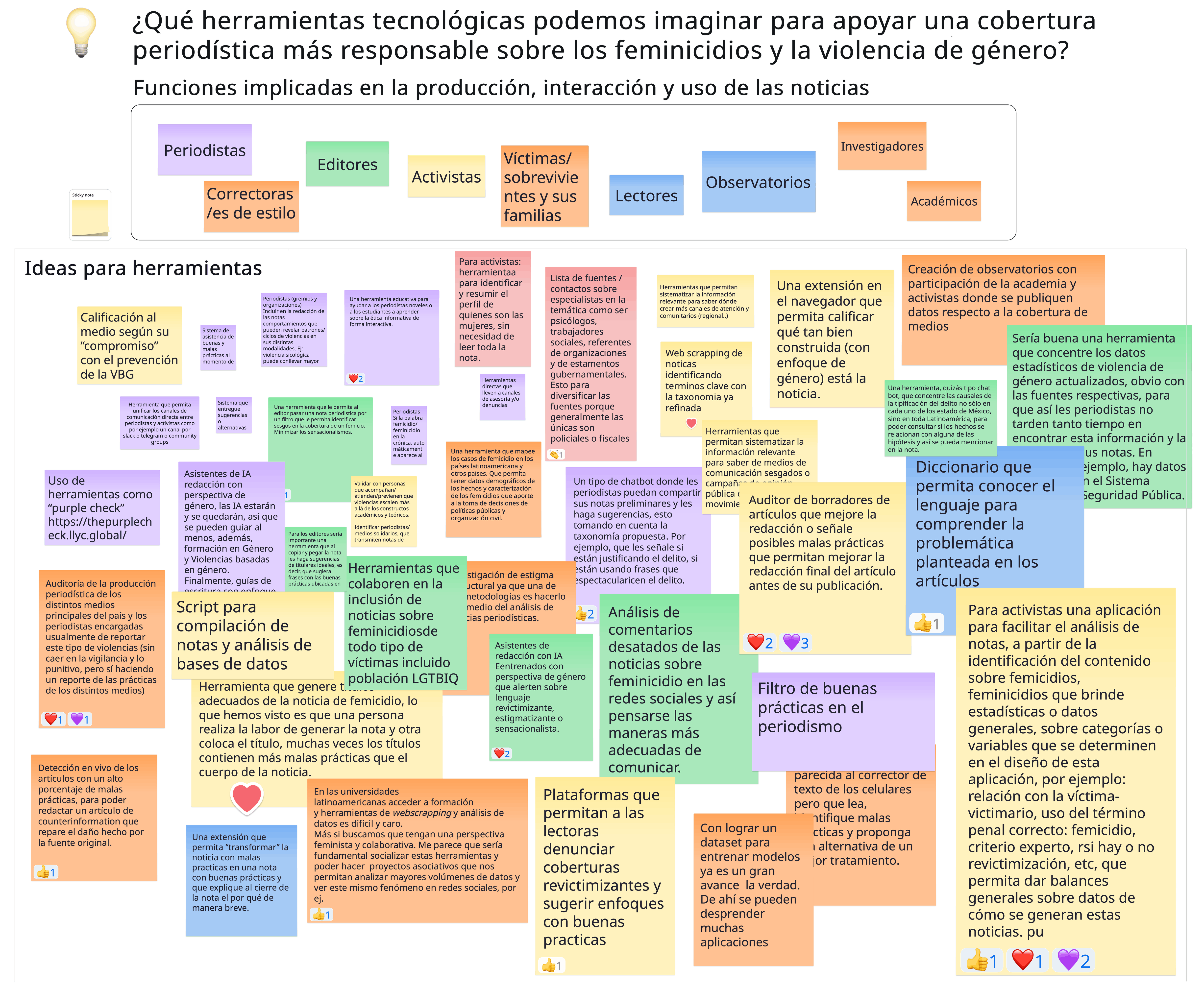}
 \caption[Collaborative sticky-note brainstorming board from the Spanish workshop (in Spanish).]{Collaborative sticky-note brainstorming board from the Spanish workshop (in Spanish). Participants ideate on how technology could support different phases of feminicide news production and readership.}
 \Description{Figure 2 shows the collaborative sticky-note brainstorming board from the Spanish workshop (in ES). Participants ideate on how technology could support different phases of feminicide news production and readership. }
 \label{fig:brainstorming}
\end{figure*}

After the workshops, we conducted open coding on breakout room transcripts, identifying emergent patterns and synthesizing insights to refine the taxonomies. Comparing proposed changes across the two languages, we found both shared and context-specific insights. For example, participants in both workshops discussed adding an overemphasis on perpetrators and a lack of victim-centered narratives to the taxonomy as a harmful practice. At the same time, participants \revise{in the English workshop highlighted the role of culturally specific categories other than ``feminicide/femicide'' (e.g., ``honor killings'' or ``dowry death''), while participants in the Spanish workshop emphasized adherence to specific legal definitions of femi(ni)cide across Latin America.} 

\subsubsection{\revise{Asynchronous} annotation}

A subset of participants (18 Spanish-speaking, 10 English-speaking) expressed interest in continued collaboration beyond the workshops, contributing to a larger annotated dataset through independent annotation work with compensation. Each participant annotated 1 to 60 articles (average=15.5), and received compensation accordingly: 
a base compensation of USD \$40 for all participants, with additional tiered compensation (USD \$50- \$150) for those who annotated larger numbers of articles ($\geqslant$ 5). Additionally, all participants are included in the acknowledgments of this paper. We expand on how we navigated decisions around compensation and recognition in Section \ref{sec:labor}.

While annotators worked on labeling individually at this stage, we designed the following mechanisms to facilitate space for ongoing collective dialogue and mutual support:
\begin{itemize}
 \item \textit{Chat groups}: \revise{We set up a messaging app} group chat for participants to ask questions, share ideas, and support each other. 
 \item \textit{Weekly discussion hours}: We organized a weekly synchronous virtual meeting space for participants to drop in and discuss articles, taxonomies, technical issues, or just get to know each other. 
 \item \textit{Communal idea and documentation boards}: We created a Padlet board as a shared documentation space for summarizing weekly meeting notes and making them accessible to those who could not attend, as well as for participants to share observations and proposed taxonomy changes.
\end{itemize}

Discussions emerging from these spaces continued to shape the dataset and taxonomy, as participants problematized category schemes and definitions, suggested ways to refine them, and continued contributing news articles from their contexts. For example, realizing that there was a lack of positive examples of responsible reporting, 
participants proactively searched and shared new articles to increase representation of constructive practices. Later, as the asynchronous annotation stage wrapped up, two participants contacted us with an interest in studying disagreements in the annotated dataset and organizing a follow-up session to discuss them and reflect on the annotation process with other interested participants. 
In this way, as the iterative process of workshops and data annotation unfolded, participants became our collaborators, contributing valuable expertise and experience to shape the dataset and future of the project.


\subsubsection{Participant Feedback}
To better understand participants' experiences in the workshops, we organized a feedback survey on their motivations for joining the project, what they found valuable or challenging about participation, and their perspectives on how their work should be acknowledged and compensated. \revise{We discuss insights from the surveys---particularly regarding the value of pedagogical spaces, community connection, and compensation---in the next section. }

\section{Reflecting on Workshops as a Methodology for AI Development}\label{sec:reflections}

Over the course of the workshops' design and implementation (roughly October 2024 - September 2025), the research team met regularly to reflect and iterate on our process. Drawing on the reflexive approach of \citet{rosner2016out}, we aimed to understand the possibilities for data and AI tools to support news coverage of gender-related violence (``practice-based'' inquiry) and reflect on the affordances of the workshop format and our research process itself (``practice-led'' inquiry). 
As we conducted each additional step of the project, members of the research team analyzed and conducted open coding of new empirical data, including workshop and discussion transcripts, brainstorming boards, and anonymous feedback forms, iteratively identifying patterns across these data. These insights, along with our own experiences as facilitators and co-organizers of the workshops, informed continual critical reflections on the challenges we faced and outcomes we observed when enacting design goals (\textit{acknowledging labor}, \textit{considering context and pluralism}, \textit{centering care}, and \textit{fostering collective dialogue}) in practice. \revise{In this section, we synthesize these insights and reflections, interweaving illustrative quotes and drawing on extended vignettes in Appendix \ref{appendix}
to ground our analysis in participants' contributions\footnote{Participants are referenced with P01-80, while facilitators (who were not also participants) are referenced with F01-09. P01-24 were English-speaking, P25-74 were Spanish-speaking, and P75-79 were Portuguese-speaking. Many participants are multilingual, but are listed according to the language group they preferred to join.}. These insights add depth to existing normative principles and highlight enduring tensions that can inform future efforts towards ``building AI systems otherwise'' \cite{ciolfifeliceDoingFeministWork2025}.}

\subsection{Pluralism within Bounded Contexts}\label{sec:context}
As we have discussed, traditional data annotation paradigms in AI development reproduce and reinforce unequal power structures between well-resourced ``creators'' and isolated ``annotators.'' The resulting artifacts tend to reflect Western, educated, industrialized, rich, and democratic (WEIRD) values \cite{Smart2024DisciplineAL} and exclude perspectives from other groups and geographies \cite{miceli2022data}. To avoid this pitfall, we were guided by the feminist principle to \textit{consider context and pluralism}.


However, as we aimed to design annotation processes to be context-specific, we grappled with the question of what ``context'' means in practice and how it relates to pluralism. 
\revise{In our initial workshop, we prioritized pluralism, bringing together perspectives across geographic and cultural contexts. At the same time, the computational requirements of AI systems typically call for standardized and consistent training data, which led us to initially pursue a unified taxonomy\,---\,i.e., a structured set of categories
with clear boundaries that could encompass reporting practices across contexts. As discussed in Section \ref{sec:intro-workshop}, however, our first workshop revealed that convening pluralistic perspectives from around the world towards this aim was both impractical and risked flattening important linguistic and cultural contexts. Rather than attempting to synthesize different perspectives through mechanisms like majority voting or
top-down decision-making by the organizers, we decided to switch subsequent workshops to be language-specific, hypothesizing that language might serve as a proxy for cultural and geographic
proximity. We hoped that these more ``bounded'' contexts would enable more culturally-sensitive versions of the taxonomy through what we characterize as \textit{tactical consensus}\,---\,a practical agreement, collectively negotiated, that meets the structural requirements of technology development for a specific context.}

In the subsequent Spanish workshop, we found that participants mostly came from Latin America\,---\,a geographic region with both significant diversity and cultural and historical commonalities. Then, in breakout rooms, we split participants into country-specific (e.g., Mexico) or region-specific (e.g., Central America) groups. We found that this configuration spurred discussions at different levels of granularity. \revise{For instance, as illustrated in the vignette from Appendix \ref{vignette1}, in the Mexico-specific breakout group, participants leveraged their shared understanding of regional issues---such as power dynamics and stigmatization of sex workers---to negotiate new categories that better scrutinize how perpetrators are humanized in media narratives.} 
In cross-regional breakout groups, \revise{participants clarified nation-specific norms to uncover deeper structural biases in reporting.}
Here, facilitating dialogue within specific geographic boundaries enabled participants and us to explore cultural nuances \revise{in relation to feminicide reporting, and productively iterate on a shared taxonomy that incorporated them}. \revise{For example, the vignette in Appendix \ref{vignette2} shows how participants explored cultural nuances---such as the stigmatizing implication of `machete blows' in Caribbean contexts versus gun violence in U.S.-influenced regions---and consequently refined taxonomy categories (``othering'') to account for how certain crime details serve as socio-economic markers. }


In contrast with the Spanish-language workshop, participants who attended the English-language workshop came from all parts of the world, and most spoke English as a second language. For this workshop, due to participants’ varied geographic locations, we assigned breakout rooms to represent larger continental regions (e.g., Asia; Europe). In these spaces, many discussions revolved around differences across contexts (e.g., categories to name lethal gender-related violence: feminicide, honor-killings, dowry deaths, etc.). Participants also brought up regional particularities relating to media production. A participant from Egypt, for instance, noted, \revise{``\textit{English-language reporting of feminicides, written by younger journalists and geared towards Western audiences, is usually more progressive than Arabic-language coverage in the region. If you only look at English, you lose the context of how it is discussed in Arabic}'' (P07).} 
However, given the vast geographic distribution of English-speaking workshop participants (from Nepal to Kenya), it was challenging for these kinds of contextual dimensions---including situated regional features, colonial histories, and media economies---to meaningfully inform \revise{practical agreements around the taxonomy} or the annotation process. 
Overall, \revise{these conversations indicated that} while English could serve as an articulating device enabling cross-cultural dialogue across diverse viewpoints, its status as a currency language enabled ``too much'' diversity, to the extent of subverting our goal to attend to cultural specificity. 

More broadly, comparing the Spanish and English-language workshops highlights the need to delimit which elements of shared experience (e.g., geography, language, profession, etc.) provide the refined contextual knowledge required for the collective annotation of a particular dataset. For our current project, because the data are news articles, we have found that lived experience grounded in geographic location is key to understanding cultural dimensions of feminicide \revise{reporting} and building \revise{tactical} consensus around \revise{which journalistic practices to include in a working taxonomy}. We found that for the primarily Latin American Spanish-speaking group, it was possible to build a working agreement while maintaining commitments to culturally-sensitive perspectives through synthesizing discussions at country- and region-specific levels of granularity. But this was not as straightforward for the more geographically diverse English-speaking group. \final{We also acknowledge that our project benefited from shared feminist commitments across participants, which reduced frictions, while other contexts may involve deeper value or power asymmetries.} We raise these underlying tensions as fruitful questions for further work on participatory data annotation to continue exploring: When is it possible to reach productive agreement while maintaining a commitment to \final{power-aware,} plural, and culturally-sensitive perspectives? 
When do disagreements instead reflect foundational differences in perspective that would be flattened through attempts at standardization? \final{We offer the concept of tactical consensus not as an invitation to sweep power differentials under the rug but to contend with these questions. The concept invites reflection on how we can create spaces where participants collectively negotiate underlying power structures with the most marginalized voices centered rather than sidelined.}
%

\subsection{\revise{Fostering} Multiple Depths of Dialogue}\label{sec:dialogue}
Typically, the role of data annotators is limited to the literal step of annotation: they are to assign a label given a classification scheme, data point(s), and instructions. Beyond this step, annotators have minimal influence in shaping how meaning is created from data \cite{Smart2024DisciplineAL}. Through our workshop \revise{methodology}, we aimed to \textit{foster collective dialogue} as a mechanism to instead involve annotators as co-contributors with valuable expertise for shaping the project. 

We found that by creating conditions for participants to engage in collective dialogue, participants naturally began addressing different aspects of the process, such as the taxonomy, labeling interface, dataset composition, and  brainstorming of possible tools. \revise{We illustrate an example of participants refining taxonomy categories in Vignette 3 in Appendix \ref{vignette3}.} \revise{As these conversations developed}, we iterated on the workshop structure to facilitate \revise{various} levels of engagement. For example, in the second round of workshops, we provided a form for participants to contribute their own articles for annotation ahead of time, and brought on some participants as facilitators for breakout sessions. We also built in time for brainstorming about the future of the project and discussing how to (not) use the data for possible AI applications. During this time, participants \revise{collectively} reflected on their values and concerns, and questioned underlying assumptions about AI’s role in their work. \revise{Building on \citet{ciolfifeliceDoingFeministWork2025}'s notion of ``building AI systems otherwise,'' we found that collective dialogue can help shift data annotation from a technical and ``neutral'' preparation step into a political and relational practice that shapes what AI systems can do.} 

\revise{At the same time, a central insight from this process is that} ``more'' (more involvement, at more stages of the data production process) is not necessarily a desirable \revise{approach} either. Participation, especially when involving a sensitive problem like gender-related violence, has significant ``material and affective demands'' \cite{dourish2020being, suresh2024participation}; not everyone may want or be able to be heavily involved in all stages of the process. For example, in our project, while 14 (English-speaking) and 30 (Spanish-speaking) participants attended the second round of workshops, only 10 (English-speaking) and 18 (Spanish-speaking) chose to continue asynchronously annotating articles. 

\revise{In this way, the methodological approach we devised over the course of the study and which we put forward to others is one that facilitates \textit{multiple depths of dialogue}. Facilitating multiple depths of dialogue in data annotation requires viewing annotation not as a single task, but as many interdependent knowledge construction processes: constructing taxonomies or label definitions, designing the labeling interface, curating data to be labeled, applying labels to data, resolving disagreements, shaping future uses of the data, and defining the annotation workflow itself. It also requires questioning the ``implicit normative assumptions'' that cast higher degrees of participation as better \cite{shirk2012public, cornwall2008unpacking}, aligning with approaches like \citet{gomez2024participation} that enable participants to choose their own degree of participation. 
This reframing brings into view a design space that expands how we might imagine and structure participation within annotation work:
How might we create opportunities for dialogue and co-creation at many key points, while giving participants the agency to determine their own levels of involvement? And how might we allow roles to shift over time, as participants navigate the demands of the work, its affective burdens, and their own evolving desires?}


\subsection{Acknowledging Labor}\label{sec:labor}

Data annotation involves intellectual, emotional, and embodied labor. \citet{miceli2022data} theorize the \textit{data-production dispositif}, whereby data work is both undervalued, fueling the demand for more and cheaper data, and largely invisible, maintaining the façade of AI as fully automated and neutral. We aimed to \textit{acknowledge the labor} of data annotation to challenge these hierarchies of power and knowledge, and make visible the skill, time, and (physical and emotional) energy required for AI development. \revise{We asked: what (political) relations would different structures of acknowledging labor enable?}

\revise{Since the early stages of the methodology design, we reflected on how participants' contributions should be adequately acknowledged and valued. While our perspective resists a hegemonic capitalist logic that reduces value to monetary terms, we also understand that material compensation is an important dimension of confronting exploitative labor relations and extractive research practices. In particular, we wanted to avoid relying solely on participants' good will, interest, and availability, which could create uneven dynamics regarding who is able to participate, and might inadvertently reinforce a self-sacrifice culture in human rights activism \cite{hernandez_cardenas_self-care_2017}, particularly for women facing a double or triple burden \cite{gupta_triple_2005}. Throughout the project, we grappled with and continually discussed how to compensate participants fairly while navigating the practical tensions of paying participants across different global contexts.}

\revise{For the first introductory workshop, we decided not to include monetary compensation for several reasons. First, this workshop primarily aimed to introduce the project to potential participants, serving as the first touchpoint of a longer, sustained partnership with a community of interested people. And second, since the recruitment call was broad and open, we received a high number (113) of registered participants---significantly more than we expected would go on to engage in the broader project---making compensation for this initial workshop logistically challenging within our institutions. During this initial workshop, however, we underscored to those who were interested in contributing further to the project that their participation and annotation work in subsequent workshops would both be acknowledged and compensated transparently. For the following workshops, participants received \revise{USD} \$60, and those who took facilitation roles received \revise{USD} \$100. }

\revise{For the self-directed annotation work, we had informed participants that they would be materially compensated but did not mention specific amounts at the start, as we continued internal discussions about how to best approach this. As discussed in Section \ref{sec:dialogue}, we sought to facilitate multiple depths of dialogue and engagement with the project. And indeed, in this stage, some participants annotated more articles than others, some spent more time on individual articles, some joined the informal meet-ups while others did not, and some shared more articles to add to the dataset. How to acknowledge and compensate these various forms of contribution? We debated whether they could or should be measured in some way (e.g., the number of annotated articles or the time spent annotating) and whether quantification would be fair -- given that the emotional and/or time investment in participation would weigh more on some participants. An important reflection for us was a desire to acknowledge various forms of contribution without simply reproducing economic rationales that privilege productivity or scale. We considered posing these questions directly to participants, but we also believed this risked burdening them with consultation on a matter they were not equipped to meaningfully influence, particularly given budgetary and administrative constraints and the logistics of international payment systems. In the end, we pursued a more pragmatic approach: we offered a base payment (USD \$40) for all those who participated in this stage and defined a tiered payment structure (an additional USD \$50 - \$150) to acknowledge participants who had annotated more articles. From our perspective, this approach balanced our desire to value all depths of engagement while recognizing participants who went above and beyond in their contributions.} 

Still, we ran into logistical challenges with paying participants across global contexts. We found a lack of supporting infrastructure or guidance within our institutions, resulting in a time-consuming process and delayed payments for some participants, as well as putting researchers in the position of advancing funds. We are also aware that across global contexts and given participants with diverse socioeconomic backgrounds and employment status, the value of a particular amount of money can vary widely, and are grappling with the extent to which this should be taken into account when determining compensation amounts. \revise{Despite these challenges, the feedback we received on the compensation process was largely positive, although we are aware that negative feedback may remain unspoken. Participants conveyed their gratitude and even surprise at the compensation process, comparing it with other projects they had taken part in as participants or researchers. One participant told us in an email that she \textit{``always had trust in the process''} and expressed joy in \textit{``being part of something bigger along with incredible feminist activist women working towards a more solidary and empathic world''} (P44). }

\revise{In addition to monetary compensation, we considered alternative means of acknowledging participants' contributions. } \revise{We explored these questions through our feedback survey with participants, which highlighted their desire for forms of recognition beyond a purely transactional, monetary model, such as educational opportunities, skill development, mentorship, and public recognition, which we are considering moving forward. }

\subsection{Centering Care}
We also aimed to recognize the emotional labor of the work through \textit{centering care}. Annotating news articles about feminicides is a sensitive and emotionally charged process that takes a toll on mental health \cite{DAFblog}. During the workshops, we found that community relationships and support helped shift data annotation from an isolated task into a collaborative and relational practice. Annotating data together fostered a sense of solidarity amongst participants, and some contributors quickly bonded over shared passion and commitment to addressing gender-related violence. For example, in one Spanish-speaking breakout room during the introductory workshop, participants became visibly angry after seeing sensationalist headlines that trivialized the crime. This emotional reaction created a sense of urgency and motivated participants to annotate articles even before the formal process began. The breakout rooms became spaces where participants transformed grief and anger into public action and meaningful critiques of the articles, described by F06 as \textit{``having fun but together being outraged''} in a facilitator debrief meeting. Participants also shared strategies for self-care and the importance of caring for the ``mind-body'' when engaging with heavy topics. For example, one participant (an academic researcher) shared their experience of integrating meditation into their daily life to cope with the emotional weight of studying feminicides and recommended the practice to others. 

As we moved into the asynchronous annotation stage, we aimed to continue cultivating community care and relationships. We encouraged participants to take breaks when engaging with heavy materials, and created spaces, including group chats and weekly video meetings, for participants to convene. Participants helped each other with questions or uncertainties in the group chat, and also came to check in with the research team during the weekly hour, sometimes to discuss specific articles, and sometimes just to get advice about their own work or chat about personal struggles. \revise{One participant later shared with us via email that \textit{``the conversations [in weekly discussions] with women who also work on gender issues were profoundly enriching''} (P44).} Importantly, the community built through the workshops provided a foundation of rapport and solidarity that extended into the asynchronous annotation stage.

\subsection{Rebalancing Epistemic Hierarchies}
The conventional conditions of data annotation reflect stark hierarchies of knowledge production. The ``high-value'' work of defining problems and creating classification schemes is distanced from the ``low-skill'' work of actually annotating data \cite{muldoon2024feeding}. This distance is maintained through geography, a lack of feedback channels, and the stratification of to whom these different forms of work are relegated. And it is reinforced through epistemic hierarchies: data requesters, for example, are often formally trained in data science, research, or software development—ways of knowing that traditionally are granted more authority and legitimacy over the experiential or embodied knowledge that annotators might bring to bear \cite{widder2024epistemic}. 

We found that through \textit{fostering collective dialogue}, feminist workshops can provide a site for capacity building and mutual knowledge sharing that supports the disruption of these stratifications. Breakout rooms became spaces for mutual learning, where participants shared their expertise on gender-related violence and journalism with each other and facilitators. Participants organically emerged as both teachers and learners, and drew upon each other’s specialized expertises. In an English-speaking breakout room, for example, participants consulted the journalist among themselves before making assertions about norms around journalistic practices. \revise{One participant reflected afterwards, ``\textit{Having a journalist in the group is useful because we might see some sentences as problematic, but they can tell us when that’s actually standard practice in newsrooms.}'' (P28)}. 
\revise{Throughout the process, facilitators also reflected on learning from participants:} for example, during the introductory workshop, while each facilitator began by demonstrating the annotation software through a pre-annotated article, participants frequently identified issues with the article \revise{--- in one case, as many as \textit{``six other things that weren't pre-annotated''} (F06) ---} that facilitators had overlooked, such as implicit victim-blaming or culturally sensitive interpretations. 

We also built in pedagogical components on AI and data production into the workshops, particularly emphasizing feminist approaches and potential applications of AI in activists’ and journalists’ daily work. This approach was inspired by Freire, who stated ``to teach is not to transfer knowledge but to create the possibilities for the production or construction of knowledge'' \cite{FreireFreedom}. To this end, our intention was to demonstrate that one does not need to have a PhD in computer science to engage critically with AI and the data economy. Indeed, feminist activists and scholars already have many tools of critical thinking and political engagement that are urgently needed in public conversations about AI. Participant feedback supported the value of learning and sharing knowledge: in the feedback survey, all participants reported ``learning about feminist AI'' as a primary reason for joining the workshops. \revise{They also valued both the pedagogical elements of the process and the space to exchange knowledge and ideas with other community members, }and all either agreed or strongly agreed on a 5-point Likert scale (average rating 4.6 out of 5) that they gained knowledge relevant to their own work.

\subsection{AI development as a Heuristic for Critical Reflection}
\revise{We initially started this study with a possibilistic view of what an automated system can contribute, informed in part by prior success with AI-based tools for activists through the DAF project and participants' interest in exploring what feminist AI can accomplish. Over the course of the project, this enthusiasm became the starting point for critical reflection.}
We intended for the AI development stage of the project to be open-ended, leaving the door open for participants to envision if and how AI should have a role in addressing media coverage of gender-related violence. In the brainstorming sections of the workshops, participants actively exchanged ideas not only around possible uses of AI, but also about their fears and concerns around AI usage. 
\revise{Despite generating a myriad of ideas on possible AI tools, participants worried about the ``homogenization of stories,'' the loss of ``humanism'' in writing, and the risk that ``AI models generate false or inaccurate information.''} 
P42 was hopeful that the very process of producing data and prototyping an AI tool could serve as an educational opportunity \revise{\textit{``to critically engage with media coverage and reporting practices, such that the AI system itself may ultimately be unnecessary.''}} 
Indeed, from the outset, we approached the AI development process as a useful heuristic device to gather community and critically reflect on broader questions around news coverage and the potential role (or refusal) of AI in activism. Capacity building and mutual knowledge sharing among the community of participants and organizers emerged as core objectives in and of themselves, separate from the creation of an AI tool. 
Reflecting on this, we ask more broadly: what would it look like for ``AI development'' processes to decenter the inevitability of AI? 


\section{Discussion and Future Work}\label{sec:discussion}


Data annotation, a crucial foundation underpinning AI, remains undervalued and frequently outsourced to precarious workers in the Global Majority. 
The work presented in this paper aligns with scholars and activist movements calling for alternative forms of data and AI production. We build on a growing body of work that aims to challenge the power, inevitability, and extractiveness of AI \cite{varon2021ai_consent, suchman2023ai_thingness, MIT_AI_Colonialism}, to reimagine a more sustainable and liberatory future for technology \cite{quijano2024feminist,DF_for_AI, ricaurte2022feminist_decolonial_ai,ciolfifeliceDoingFeministWork2025}. We contribute a concrete case study of collaborative data annotation workshops designed to consider context and pluralism, center care, acknowledge labor, and foster collective dialogue, \revise{highlighting the relational and political engagements made possible through the workshop space. In this section, we expand on insights, tensions, and lessons learned from our process, and their implications for future work. }



\revise{Through enacting feminist and participatory principles in practice, our case study surfaces insights that both add depth to and complicate these principles. 
For instance, while acknowledging labor has been discussed broadly in the critical HCI community, our reflections highlight that enacting this principle requires us to think beyond compensation from a purely economic logic. Similarly, while participatory approaches to AI have been increasingly adopted and studied \cite{delgado2023participatory}, our work points to the need for \textit{multiple depths of dialogue} that both expand opportunities for participants to shape meaningful aspects of the process---such as data curation, taxonomy construction, or tool ideation---and give participants the agency to choose how and to what extent they engage. }
By carefully shifting between owning and distributing the power differential in the researcher-participant hierarchy---taking responsibility for decision-making when appropriate while acknowledging annotators as co-creators, rather than as passive laborers, and collaborating over shared goals in the long term---these practices help move beyond a monolithic understanding of ``participation'' towards \textit{rebalancing traditional epistemic hierarchies}. And, through considering context and pluralism in practice, our reflections emphasize the importance of ``bounding'' context to facilitate deeper dialogue across pluralistic perspectives and work towards a \textit{\revise{tactical} consensus}.

As we continue to reflect on our commitment to pluralistic perspectives, however, 
\revise{we must also interrogate the need for consensus in the first place.}
Even within a shared context, there is still richness in different lived experiences and diverse viewpoints. 
The dominant logic and technical requirements of AI development, however, often push us towards defining clear and consistent ``ground truth'' labels. Shifting these norms may require iteratively deconstructing the notion of a singular ``ground truth'' while also developing methods that do not necessitate it. 
We point towards opportunities for annotation software and processes to preserve and documenting meaningful disagreements---for example, through capturing annotators' rationales or confidence levels\revise{---or working towards building tactical consensuses, where participants agree on the most politically useful annotation strategy to achieve a specific goal}. In parallel, we highlight the promise of lines of work in AI that aim to incorporate subjectivity and disagreements as feature sets into models \cite{Basile2021TowardAP}. 
\revise{Moving forward, we are actively experimenting with and evaluating technical approaches using our dataset, and we invite the broader research community to similarly ground pluralistic AI development in rich, real-world case studies.}


\revise{We can also turn to examining tensions that emerge in the} process of organizing participatory projects. In our work, prioritizing reflexivity, iteration, and an open-ended process was key to building lasting relationships and meaningfully collaborating with participants. Yet, this approach entails a significant shift in thinking about the time commitment and organizational load of a project.
A task that could be unilaterally defined by a small team of researchers, outsourced through a platform, and ``completed'' in a week might instead take months---necessitating care and effort at odds with the prioritization of efficiency and standardization in ``scale thinking'' \cite{hanna2020scaleprovocationsresistancesscale}. 
\revise{For example, the decision to move from a multilingual format to separate language workshops duplicated the organizational load on the research team, leading to a slower timeline for data collection and annotation (and we are still into the process of discussing a second Portuguese workshop, as we review our capacity to replicate the process for a third context). }We encourage future work to continue experimenting with practices of care and alternative collaboration schemes that resist returning to conventional \revise{drives for efficiency and scale. }

\revise{Finally, while our process began with a question around curating datasets for AI development, this inquiry ultimately served as a heuristic device to critically engage with media reporting of feminicide and broader considerations around the role of AI in this space. }
These complicated negotiations and debates often open more questions rather than a neat \revise{linear progression}. Our open-ended process, in turn, allowed us to continually challenge our assumptions and iteratively adjust our goals and process, including questioning whether AI should be used at all. We argue that such sustained \revise{community-building and collective dialogue} is essential to disrupting the extractive and colonial conventions of data and AI work, and to reimagining these processes as relational and political spaces for understanding differences, enacting care, and building solidarity across shared struggles.

\section{Conclusion}
This paper presents a case study of building a multilingual, collaboratively annotated dataset of news coverage of feminicide. It \revise{makes a methodological contribution} to critical HCI and AI scholarship by moving beyond articulating conceptual principles to implementing feminist and participatory data annotation practices. We demonstrate how data annotation workshops rooted in feminist commitments can engage pluralistic perspectives, foster collective dialogue, and disrupt knowledge hierarchies.
In reflecting on the tensions of translating normative principles to practice, 
\revise{we deepen existing normative principles and articulate specific methodological strategies that can help guide future efforts to build AI otherwise.} Through this work, we contribute to an alternative political and economic vision for data and AI work that elevates agency, relationality, care, and solidarity. 


\begin{acks}
We would like to thank the Pembroke Center at Brown University for supporting this project. We are grateful to our workshop facilitators Mariel Garcia-Montes and Alyssa Lanter, and to Claudia Laudano, Daniela Paz Moyano Dávila, Florencia Pagola, and Maria Gargiulo, who generously contributed as both facilitators and participants. Finally, this work would not have been possible without the participation and insights of our workshop contributors, to whom we are deeply grateful:

Aimee Zambrano Ortiz,
Ana Paula Galindo,
Ana Rita Argüello Miranda,
Andrea Olga Lague,
Angélica Barreno,
Anna Kapushenko,
Anna Sochorova,
Anthony Lee Medina Nieves,
Azalea Reyes Aguilar,
Camila Bautista Torres,
Carolyn Carina Casado Fontalvo,
Cathy Otten,
Claudia Laudano,
Clemi Niño,
Daniela Paz Moyano Dávila,
Diana Ferreiro,
Dreyf Assis Gonçalves,
Dunja Bonacci Skenderović,
Edda Ileana López Serrano,
Edda López Serrano,
Ekemeren Kaneng Gbadamosi-Ladoja,
Elena Iacovou,
Fabiola Gutiérrez,
Florencia Pagola,
Gabriela Coronado Téllez,
Gema Villela,
Gihomara Aritizábal Morales,
Hana Wael Aly Mohamed Elrashidy,
Helen de Souza,
Humaira Ratu Nugraha,
Empar,
Janvi Singh,
Jordana Luz Queiroz Nahsan,
Julie Rajan,
Karla Dayana Castro Polanía,
Kristina Zakurdaeva,
Laura nudm roma,
Lizbeth Escudero,
Luz Alejandra Pedreros Sierra,
Manuela Cau,
Maria Arboleda,
María Asunción Collante Jara,
Maria Gargiulo,
Mariana Aldrete,
Mariana Gazcon Nuñez,
Mariana Saldarriaga,
Marina Garcia-Vasquez,
Marina Mieres,
Marisol Anzo-Escobar,
Mercedes Altuna,
Mia Della Vita,
Monica Lopez,
Mónica Pellegrino,
Monica Zocconali,
Monti,
Nadia Donoso Sánchez,
Nancy Pretto,
Natacha Xavier,
Natalia Alcocer Rosas,
Nataša Vajagić,
Natia Johana Lombardo Da Luz,
Ololade Ajayi,
Paul Alexander Parra Ñiquen,
Rivera Quishpe Kerly Aracely,
Rocio Sarasola,
Samantha Páez Guzmán,
Savia Hasanova,
Silvina María Luz Molina,
Stephanie Cobo Maturana,
Susana Barradas,
Sushmita Panda,
Tatiana Walk-Morris,
Teresa Herrera,
Teresa María Herrera Sormano,
Tufan Neupane,
Valentina Agredo Sanín,
Valeria Acosta Isaza and
Virginia Cavallaro.

\end{acks}





\bibliographystyle{ACM-Reference-Format}
\bibliography{references}

\appendix
\section{Appendix A: Workshop Vignettes}\label{appendix}

\subsection{\textbf{Vignette 1: Discussing stigmatization, power imbalance, and perpetrator representation in Mexico}}\label{vignette1}

\noindent\textbf{Context}: In the Mexico-focused breakout room in the Spanish workshop, participants discussed a news article that depicted a stigmatized victim against a socially protected perpetrator. By collectively analyzing regionally specific issues, such as the stigmatization of sex work and nacro-trafficking, the group argued that the draft taxonomy was insufficient for capturing how media narratives portray powerful aggressors. This conversation took place in Spanish, and is translated to English below. 


\smallskip
\hrule
\smallskip

\begin{quote}
\textbf{P36}: In my database, if you search for the aggressor, there’s a specific category called client, because in the Mexican context, a lot of cases involve sex workers. [...] Trans women especially are heavily stigmatized. I’m thinking here maybe stigmatization by occupation needs to become its own category.

\textbf{P45}: Yes, and what really stands out in femicide reporting here is the imbalance: everything is about the victim, and there’s almost no information on the perpetrator. [...] 
I think good coverage is not just ``respectful toward the victim.'' It also means not glorifying or demonizing the perpetrator. Both extremes distort the story, which would reinforce class and moral hierarchies that are very prominent in Mexico.

\textbf{P36}: And this imbalanced coverage ends up working against the victims in most cases. In this article, what sticks is how they insist that she was an escort---like they try to soften it with ``she went into that because of the pandemic,'' but as the piece goes on, her ``moral quality'' keeps getting downgraded. Meanwhile, his prestige keeps going up: he’s first just a client,  and then suddenly he’s a ``banker.'' In our context, \textit{banquero} sounds like someone powerful, even though we don’t know if he owns a bank or just works at one. His prestige grows as hers declines. 

\textbf{P25}: Listening to all of this, I agree that we should add something to the taxonomy about the perpetrator: what is said about him, what he does for a living. Not just his relationship to the victim, but also details that often are not relevant. And we can read this through class, through sex work, and also through the Mexican context of narcotrafficking. Articles often mention a person’s relationship to crime and use that to justify what happened. So I think we need more annotations around the victimario, because it changes the narrative if he is a client, a partner, or a family member.

\textbf{P45}: I was thinking of the femicide count that Serene Young does in Nicaragua: she codes the perpetrator’s profession. That’s useful to break the ``otherness,'' to show that it is not only people from the peripheries or linked to narcotrafficking; perpetrators can also be professionals, students, university students, and so on. In the Mexican context, it would be interesting to have a tag for when the crime is criminalized or justified through a supposed link to narcotrafficking. María Salguero calls these \textit{feminicidios territoriales}, because drug violence is so tied to territory.

\end{quote}

\subsection{\textbf{Vignette 2: Understanding how weapons contribute to othering in different regions}}\label{vignette2}

\textbf{Context}: In the Caribbean (cross-regional) breakout room of the Spanish workshop, participants annotated a Cuban news article about the killing of a young woman described as ``a 17-year-old, mother of a girl and resident in a batey.'' This conversation took place in Spanish, and is translated to English below. 

\smallskip
\hrule
\smallskip

\textbf{P32}: Sorry, I have a question. Do you know what a \textit{batey} is?

\textbf{P34}: Yes, it’s a Caribbean term. [...] In a typically precarious community, you have the settlement and then the batey is like a communal space in the center, usually surrounded by houses and other structures. It’s a community space, but batey can also refer to the whole community settled around it, a small town.

\textit{With this shared understanding of ``batey'' as a precarious communal space, the group then turned to another key sentence in the same article:}

\textbf{P32}: I have a question about something that comes up a lot in this article. What do you think about the way the article describes how the woman was killed? For example, here it says that Tamayo was killed… allegedly by her ex-husband, identified as Jordan, “who inflicted several machete blows on her.” Do you see that as a bad practice, or not really relevant? Did you mark it in any way?

\textbf{P34}: From a non-violent perspective, if we’re using the news as a tool not so much to educate but to raise awareness, I don’t think that level of detail is relevant. They could have just said that she was “wounded with a bladed weapon,” and people would understand that it wasn’t a gun. […] From the point of view of accompanying surviving victims, I know that a wound from a bladed weapon implies a lot of rage, because it requires being physically close to the victim. Unlike a firearm, which you can use from a distance. So when I reconstruct the feminicide, just knowing that it was a bladed weapon and that it happened in her home already lets me infer harassment, surveillance, invasion of her space over time – a crime of proximity. But that doesn’t mean the media have to spell out ``several machete blows.''

\textbf{P28}: Also, when these cases happen in impoverished communities, mentioning the specific weapon is a way of constructing an ``other.'' When it’s a gun – which also happens – there’s a class layer there, and it often doesn’t get mentioned in the same way as weapons that are already associated with certain sectors of the population. [...]
In theory, knowing the location of the crime is important because it allows us to map where things are happening, [...] but in reality, they’re contributing to the stigma of that community. They name a poor neighborhood and immediately the comments say things like ``everyone there is violent,'' ``the police need to intervene,'' and so on.

\textbf{P32}: I had annotated the machete part as ``sensationalism,'' but what you’re saying makes me see it also as ``othering.'' That detail is not only morbid; it also marks the community. In Chile we’ve had very high-profile feminicides in upper-class circles, and the way those are reported is completely different. The rich perpetrator’s case gets full coverage of the trial and a very dignified tone. When the victim is from a lower social stratum, the coverage becomes much more sensationalist. So, here too, the machete in a poor area feels like part of that same dynamic.

\textbf{P28}: I agree, and there’s another layer in Puerto Rico’s context. Here, following U.S.-style policies, access to gun licenses was dramatically relaxed. We went from about 60,000 licenses to around 300,000, and now people as young as 18 can get one. It’s very easy to obtain a firearm, and because we’re also a corridor for drug trafficking, it’s common for people to have guns even if they don’t show them. According to the Observatory for Gender Equity, almost 100\% of feminicides recorded in 2023 and 2024 were committed with firearms. So insisting on ``several machete blows'' in a poor batey in another country isn’t neutral; it says something about how violence in those communities is imagined.

\textbf{P32}: That is really interesting. I had been tagging this kind of phrase only as ``sensationalism,'' and using ``othering'' mostly for more explicit culturalising language. But hearing both of you I now see that, in this case, the ``other'' is being constructed by linking a specific weapon to a named, precarious neighbourhood. It is not just a detail about the crime scene; it defines a certain place and its inhabitants as a particular kind of violent subject. Our definition of ``othering'' in the taxonomy should incorporate that.

\subsection{\textbf{Vignette 3: Navigating tension between evidence of abuse and excessive harmful details}}\label{vignette3}

\textbf{Context}: In the European breakout room of the English workshop, participants discussed a news article reporting on a femicide trial. The group debated whether the inclusion of graphic forensic details constituted ``sensationalism'' (a harmful practice) or necessary evidence of a pattern of abuse (a constructive practice). This exchange illustrates how participants identified a gap in the existing taxonomy, i.e., the distinction between the quality of information and the quantity of information, and negotiated a new category.


\smallskip
\hrule
\smallskip

\begin{quote}
\textbf{P11}: I think we should add a label when the article goes into too much detail of the murder. [...] For instance, here we can clearly see paragraph after paragraph just describing... how the murders had happened. I don't know if anyone also thought it's a bit too much. 

\textbf{F08}: You're right... The [existing label] "overexposing details irrelevant to the case" is details about the victim's life, not about the case. And there is a missing label for that. 

\textbf{P01}: [Regarding the second paragraph], they are actually recreating the whole case, how he killed her and how long the knife was. [...] For me, it's a bit of sensationalism to repeat what happened. [...] I don't know if all these details are really necessary in the article. 

\textbf{P11}: It goes back to what I said about the previous article, that there are too much details of the actual crime that are not really necessary. [...] But one good thing that I can take away from this article is that it also exposes the pattern of him being violent with his partners. Even though it kind of goes also in a bit too much detail, it still shows the pattern... 

\textbf{P01}: Actually, [I have a] love and hate relationship between researcher and articles like this. [...] These things like, that he was following her, that he was threatening her... was actually the things that helped me put the puzzle of the coercive control that he was subjecting her under. So in a sense, I agree, it's good to know that there was a pattern of abuse... But then again, too much of it... can be toned down. 

\textbf{F08}: It sounds like there is something about too much information that is making us feel like the ``sensationalism'' label is not enough. Sensationalism tells us something about the quality of the information, but the excess of information is talking about quantity. It seems that we need both.
\end{quote}
\end{document}